\documentclass{article}
\usepackage{graphicx}
\usepackage{amsmath}
\usepackage{amsfonts}
\usepackage{amssymb}
\begin{document}

\title{Quantum Basis of Lorentz Symmetry }
\author{Paul Korbel\\Institute of Physics, \\University of Technology, ul. Podchorazych 1,\\30-084 Cracow, Poland}
\maketitle
\begin{abstract}
An unconventional outlook on relationship between the quantum mechanics and
special relativity is proposed. We show that the two fundamental postulates of
quantum mechanics of Planck and de Broglie combined with the \textit{idea of
comparison scale} (explained in the paper), are enough to introduce
relativistic description. We argue that Lorentz group is the symmetry group of
quantum, \textit{preferred frame} description. We indicate that the departure
from the orthodox relativity postulate allows us, in easy way, to make special
relativity and quantum mechanics indivisible whole.
\end{abstract}

\noindent\newpage

\section{Introduction}

The principle of equivalence of inertial frames basing on the Galilean
transformation provides the foundations of classical, i.e. Newtonian physics.
The classical physics is complete, it agrees well with our intuition and
treats the time as the absolute quantity. This makes that classical approach
turns out to be the natural and well-understood. However, the research on
electricity and magnetism, which finally led Maxwell in 1864 to his know
equations, and next the observations of Michelson and Morley in 1887, which
excluded the possibility of ether existence, have revealed some new physical
phenomena, the interpretation of which was very difficult in spirit of
Newtonian physics. Note, that derivation of Maxwell equations was possible not
only because of the knowledge of the properties of the electric and magnetic
fields and mutual correlations between them, but also due to the assertion
that the light always propagates at a \textit{finite} speed $c$. However, the
most surprising observation, approved just by Michelson and Morley, was that
the speed of light does not depend on the state of motion of the emitting
body. This observation has limited the range of acceptability of the theories
invariant with respect to the Galilean transformation. Consequently, the
natural need was to build up a new dynamical theory for which the observation
of Michelson and Morley would be intrinsically incorporated. Such theory was
formulated by Einstein in 1905, who imposed the physical meaning on the
Lorentz transformation. According to his interpretation the time is no longer
absolute quantity, so that ``we cannot attach any \textit{absolute}
signification to the concept of simultaneity...'' \cite{Ein1}.

At the beginning of last century, i.e. at the time when the quantum mechanics
had come into being, there appeared a new sort of physical observations
needing a departure from the classical treatment too. It is known that early
attempts of construction of wave equation for material particle in much degree
were inspired of mathematical structure of Maxwell equations. This led
Schr\"{o}dinger and others \cite{KG} to discover the relativistic scalar wave
equation, usually called the Klein-Gordon equation. However, it turned out
that this relativistic equation provided the values for the spectrum of
hydrogen atom that were not fully correct compared to the experimental data.
The source of this discrepancy, as noticed by Schr\"{o}dinger, was the lack of
electron spin in calculations being done \cite{Dir0}. Schr\"{o}dinger improved
then his results by substituting the relativistic equation for its
non-relativistic approximation (known just as the Schr\"{o}dinger equation).
This allowed him to introduce ``by hand'' the spin degrees of freedom and
correct the earlier results.

Nevertheless, within the framework of non-relativistic approach there is no
self-consistent way to introduce spin degrees of freedom into wave equations.
On the other hand, only non-relativistic approach allows us to treat the time
as an absolute quantity. In consequence on finds a sharp distinction between
the quantum mechanics and relativistic quantum field theory. Thus, the
prevailing conviction is that special relativity and quantum mechanics are the
separate realms that cannot be combined in a simple manner, if possible at all.

At the recent stage of development in physics, i.e. after hundred years later
the quantum mechanics and special relativity were discovered, it is hard to
expect that any further progress of currently discussed issues will solve the
problem of separation of relativity from quantum mechanics. Therefore, perhaps
the most plausible way to overcome this difficulty is a flashback overview
upon the old ides rather then the pursuit of new ones. In particular it is
worthwhile to reexamine the postulate of relativity of motion in the context
of quantum measurement, as well as, the meaning of time in the light of
fundamental postulates of quantum mechanics.

The purpose of this paper is to show that similarly like the Galilean group
constitutes the foundations of classical physics, the Lorentz group
constitutes the foundations of quantum mechanics. In other words, we prove
that spacial relativity is integral part of quantum mechanics. In particular,
it is shown that the description based on Lorentz symmetry does not exclude
the absolute time meaning. Nevertheless, a thorough refinement of
interpretative foundations of special relativity is needed . This work tackles
these issues in the most elementary way. The structure of the paper is the following:

In \textbf{Section 2}, we put and reexamine the Einstein relativity postulate
in the light of realistic quantum experiment. It is argued that each quantum
observation is the \textit{preferred frame} observation, where the role of the
preferred frame plays the rest frame of observer.

In \textbf{Section 3}, we indicate that it is important to distinguish two
time meanings. The first, classical meaning concerns the time as the measure
of pace of observed changes. The other, quantum meaning, concerns the time as
the energy measure in sense of inverse time units. We show that the
Schr\"{o}dinger quantum mechanics does not differentiate between the both time
meanings, however, the relativistic quantum mechanics does. Furthermore, we
also indicate that the source of Lorentz symmetry resides in freedom of choice
of\textit{\ comparison scale, }i.e. at the \textit{scale} which is imposed on
two physical quantities given in different physical units, like the energy and
momentum or the distance and time. The symmetry that corresponds to this
freedom of choice is called the \textit{scaling symmetry} and it is described
by \textit{transformation of scaling}.

Presented in \textbf{Section 4, }$1+1$ dimensional (gauge) analysis concerns
both: the space-time and momentum space description. The concept of
\textit{photonic frame}, where the \textit{scaling transformation} has
diagonal form, is introduced. We end up this section with an open question
about photon states description in \textit{photonic frame}.

In \textbf{Section 5}, we consider a \textit{photonic frame} and another one,
which is rotated around the origin of initial \textit{photonic} frame about
$45^{0}.$ This new frame turns out to be the Minkowski frame and thus provides
the proper (i.e. unambiguous) description of photon states. Additionally, one
finds that \textit{scaling transformation} of \textit{photonic frame}
corresponds to the Lorentz transformation of Minkowski frame.

In \textbf{Section 6} we still continue our $1+1$ dimensional (gauge) analysis
to show that state vector of relativistic particle with mass can be composed
of two massless (i.e. photon) state vectors. This reveals a composite
structure of massive state and points out to its space-time extensions.

Finally, in \textbf{Section 7}, we generalize our discussion onto entire $3+1
$ dimensional space-time. The quantum basis of space-time concept is
disclosed. Indeed, it is shown that the \textit{idea of scaling} constitutes
the core of Lorentz group symmetry. Next, we discuss the issue of kinematical
meaning of Lorentz formulas and provide plausible quantum-mechanical
explanation of, so-called, time dilatation effect.

\section{Preferred frame in classical and quantum approach}

The first relativity principle says \cite{Ein1} ``the laws by which the states
of physical systems undergo change are not affected, whether these changes of
state be referred to the one or the other of two systems of coordinates in
uniform translatory motion''. This principle, first of all, expresses our
faith in universal character of nature laws. However, our classical perception
has led also to the classical conclusion that there is a transformation group
that converts measurements made by one inertial observer to measurements made
by another. \ Note, that this ``principle-conclusion'' applied to quantum
world observations cannot be directly verified by any experimental technique
and become the main source of difficulties with relativity interpretation.

To explain this, let us consider first the case of single observer and ask: is
it possible for him to measure the speed of light or any other physical
quantity outside his own rest frame? According to Bell \cite{Bell2} ``the only
`observer' which is essential in orthodox practical quantum theory is the
inanimate apparatus which amplifies microscopic events to macroscopic
consequences''. However, each `observer' may observe only in `his' own rest
frame and no other possibility exists. The conclusion is that observer rest
frame, or in other words, the laboratory frame, always is the
\textit{preferred frame} where the nature laws can be discovered and described.

Next, let us consider the case of two observers being in relative motion and
assume that they measure the same physical quantity. To verify in physical way
the first relativity postulate, the observers, beside the measuring quantity,
have also to measure their mutual space-time positions. But this in turn means
that the observers have to be coupled. Since the above
``principle-conclusion'' refers only to independent observers it cannot be
directly verified on experimental way.

Of course, the same physical phenomena may be observed by independent
observers placed at different inertial frames, and we know that their
theoretical predictions resulting from the same equations (although
established for different boundary conditions) agree. However, this is just
the case when the\textit{\ relativity of inertial frames} manifests itself by
the fact that \textit{all (inertial) observers that make measurements in the
same conditions obtain the same experimental results}. In consequence, the
world seen by each of them looks the same. In other words, the relativity
principle reflects the most basic property of physical observation, namely,
its ability to reproduction.

Without doubt Einstein relativity postulate is a result of replacement of the
Galilean group with the Lorentz one in the Newton world. Just such approach is
a real source of pile up difficulties encountered with any attempt of
unification of special relativity and quantum mechanics. Simple arguments
given in the paper indicates that origin of Lorentz symmetry is rather quantum
then classical. Note, that quantum measurement, in contrary to the classical
one, cannot avoid a meaningful influence of measuring apparatus on the final
results. On the other hand, such influence must accompany each
\textit{preferred frame }observation\textit{\ }and thus must be
\textit{preferred frame} dependent. So, if Lorentz symmetry reflects indeed a
feature of description of the quantum world (what is going to shown clearly),
it must concern also the \textit{preferred frame} description.

\section{Two meanings of time and photon dispersion relation}

One may say that pure classical approach defines the time as the measure of
pace of observed changes. We will say that such time definition determines the
\textit{vital-time }meaning. The time evolution of system should then follow a
\textit{vital-time }description\textit{.} However, the quantum mechanics
indicates also another time meaning. Namely, due to the postulate of Planck,
generalized later by Einstein, the time may be used as the energy measure by
means of inverse time units. According to Planck postulate, if quantum state
is characterized by the wave of period $T$ , its energy
\begin{equation}
\mathcal{E}=\frac{h}{T},\label{en}%
\end{equation}
where $h$ is the Planck constant. Note, that eq. (\ref{en}) itself does not
provide any kinematical meaning of interval $T$, unless this time interval
refers to physical process which kinematics is well established. Nevertheless,
the very equation (\ref{en})\ set up the clear relationship between the energy
and time interval which devoid of kinematical meaning will be called further
the \textit{frozen-time }interval\textit{.} Thus, the \textit{frozen} meaning
of time simply means the energy measure in sense of eq. (\ref{en}).

It is commonly thought, of course, that there is no distinction between the
\textit{frozen} and \textit{vital} time meanings. Indeed, in Schr\"{o}dinger
approach there is no separation between both time meanings but the reason for
that is quite simple. The Schr\"{o}dinger quantum mechanics involves the
classical dispersion relation
\begin{equation}
\mathcal{E}=\frac{p^{2}}{2m}=\frac{1}{2}mw^{2},\label{klas}%
\end{equation}
where the energy of particle with mass $m$ is expressed by means of particle
momentum $p$, or particle velocity $w,$ where the latter is just the parameter
of kinematical description. Thus, the equivalence of energy formulas
(\ref{en})\ and (\ref{klas}) makes that there is no ambiguity in understanding
of time meaning. The \textit{frozen} and \textit{vital} meaning of time now,
is the same.

Many textbooks start with dispersion relation (\ref{klas}) to construct next
the Schr\"{o}dinger equation. Such approach, in fact, marks out the way which
Newtonian physics becomes the quantum mechanics. However, if there exists a
more fundamental way to introduce quantum description, one would expect that,
in particular, such approach itself should provide the form of dispersion
relation for free particle. The aim of this section is to indicate the basis
of such approach. First, we discuss the idea of energy-momentum
\textit{comparison scale} and derive the photon dispersion relation. This will
allow us to show later that postulates of Planck and de Broglie combined with
the idea of \textit{comparison} \textit{scale} are enough to derive dispersion
relation for relativistic particle with mass. Presented approach, however,
makes a clear distinction between the \textit{vital} and \textit{frozen} time meanings.

\subsection{Energy-momentum comparison scale}

The postulate of Planck relates particle energy $\mathcal{E}$ to the wave
period $T.$ In similar way the postulate of de Broglie relates the particle
momentum $\Pi$ to the wavelength $\lambda$, where
\begin{equation}
\Pi=\frac{h}{\lambda}.\label{mom}%
\end{equation}
Thus, in simply quantum-mechanical manner these two postulates express a basic
feature of quantum physics known as particle-wave duality. However, the issue
that has been not realized so far is that these two postulates also provide a
framework for relativistic description. To disclose this connection, first we
introduce a concept of \textit{comparison scale.}

Let us consider the most simple dispersion relation where particle energy
$\mathcal{E}$ and momentum $\Pi$ are assumed to be proportional quantities,
i.e. when
\begin{equation}
\mathcal{E}\sim\Pi.\label{enp}%
\end{equation}
Due to the postulates of Planck and de Broglie the same can be expressed with
the aid of reciprocal quantities, i.e. the wave period $T$ and wavelength
$\lambda$, namely
\begin{equation}
T\sim\lambda.\label{ent}%
\end{equation}
However, the momentum and energy, similarly like the distance and time, are
the two quantities given in different physical units. Thus, to compare them in
direct way (and thus to replace above proportions with corresponding to them
equalities) one needs to introduce a dimensional factor of velocity $v$, and
thus to set the \textit{comparison scale }for the momentum and energy and/or
the distance and time.

Without loosing generality, as well as, to simplify the discussion, now it is
enough to limit the considerations to two dimensional (i.e. $1+1$) space. Let
us then consider a linear dispersion relation in the form
\begin{equation}
\frac{\mathcal{E}}{v}=\sigma\Pi_{1},\label{en22}%
\end{equation}
where $\Pi_{1}$ is the value of particle momentum in given frame, and
$\sigma=\pm1.$ Since the value of $\Pi_{1}$ may be either positive or negative
the introduction of $\sigma$ allows us to avoid negative energy values.
Nevertheless, a new possible way of quantum-mechanical interpretation of
negative energy states will be discussed separately elsewhere.

In terms of reciprocal quantities eq. (\ref{en22}) takes the form
\begin{equation}
vT=\sigma\lambda_{1}.\label{en33}%
\end{equation}
The velocity $v$, which fixes the \textit{comparison scale,} is settled, but,
in fact, is a free parameter. Thus, the feature of the way which equations
(\ref{en22}) and (\ref{en33}) were introduced, is the \textit{freedom of
choice of comparison scale }$v$, or in the other words, the \textit{freedom of
scaling}. On the other hand, as long as the velocity $v$ does not refer to
particle kinematics eqs. (\ref{en22}) and (\ref{en33}) have no kinematical
meaning too.

To impose kinematical meaning on dispersion relation (\ref{en22}) or
(\ref{en33}), let us assume that the system we describe consists of particles
that always propagate at a constat velocity $c,$ no matter how big their
energies (or momenta) are. In a context of quantum field theory, actually, it
is even better to say that we assume the existence of physical system which
quasiparticle excitations (in a given medium) always propagate at the same
velocity $c.$ Note, that such idealized system, if only exist, might serve as
the reference one to provide a general way of energy \textit{mapping}
\begin{equation}
\mathcal{E}\rightarrow\frac{\mathcal{E}}{c}\equiv\Pi_{0},\label{map1}%
\end{equation}
equivalent to the reciprocal time mapping
\begin{equation}
T\rightarrow\Delta\chi_{0}\equiv cT.\label{map2}%
\end{equation}
Of course (and fortunately), such idealized system exists and is well-known,
so it will be called simply the \textit{photon system.} Thus for $v=c,$ eqs.
(\ref{en22})\ and (\ref{en33}) respectively take the form
\begin{equation}
\Pi_{0}=\sigma\Pi_{1},\label{en4}%
\end{equation}
and
\begin{equation}
\Delta\chi_{0}=\sigma\Delta\chi_{1},\label{en44}%
\end{equation}
where $\lambda_{1}$ in (\ref{en33}) was substituted for $\Delta\chi_{1}$ for
further convenience, Since the algebraic structure of \textit{photon} fields
is of no importance now, we shall call the ``\textit{photon }fields'' all for
which the dispersion relation (\ref{en4}) holds. Note, that eqs. (\ref{en4})
and (\ref{en44}) fulfill also more general conditions
\begin{equation}
\left(  \Pi_{0}\right)  ^{2}-\left(  \Pi_{1}\right)  ^{2}=0\text{ \ and
\ }\left(  \Delta\chi_{0}\right)  ^{2}-\left(  \Delta\chi_{1}\right)
^{2}=0.\label{int1}%
\end{equation}

\subsection{The freedom of choice of comparison scale}

The condition $v=c,$\ which has imposed the kinematical (\textit{vital})
meaning on time, also has fixed the \textit{comparison scale.} So, does it
mean that initial \textit{freedom of scaling} has been lost? In the case of
practical calculations, i.e. when the condition $v=c$ is used to evaluate the
physical quantities, the answer is yes! However, the answer is no if one uses
only the notation which does not refer to any special value of $v.$ This is
just the covariant notation. The arguments given in the paper are enough to
show that the source of covariant description is the \textit{freedom of choice
of comparison scale.}

Indeed, let us express the velocity $v$ by means of the preferred velocity $c
$ and some real scaling factor $\eta$ as
\begin{equation}
v=c\eta^{2}.\label{v1}%
\end{equation}
If one substitute velocity $v$ for $c\eta^{2}$ \ in (\ref{en22}), equations
(\ref{en4}) and (\ref{en44}) transform into the new ones
\begin{equation}
\Pi_{0}^{\prime}=\sigma\Pi_{1}^{\prime}\text{ \ \ and \ \ }\Delta\chi
_{0}^{\prime}=\sigma\Delta\chi_{1}^{\prime},\label{en55}%
\end{equation}
where\ \ $\Pi_{0}^{\prime}=\frac{1}{\eta}\Pi_{0}$ \ and \ $\Pi_{1}^{\prime
}=\eta\Pi_{1},$ and thus $\Delta\chi_{0}^{\prime}=\eta\Delta\chi_{0}$ and
\ $\Delta\chi_{1}^{\prime}=\frac{1}{\eta}\Delta\chi_{1}.$ It must occur as
well
\begin{equation}
\left(  \Pi_{0}^{\prime}\right)  ^{2}-\left(  \Pi_{1}^{\prime}\right)
^{2}=0\text{ \ and \ }\left(  \Delta\chi_{0}^{\prime}\right)  ^{2}-\left(
\Delta\chi_{1}^{\prime}\right)  ^{2}=0.\label{int2}%
\end{equation}
So, the freedom of choice of comparison scale $v$ might be alternatively
expressed through the preferred velocity $c$ and scaling factor $\eta.$

\section{Scaling transformation of \textit{photonic frame}}

In this section we address a problem of symmetry induced by the
\textit{freedom of choice of comparison scale. }Since the energy and momentum
are, in general, the two independent parameters, dynamical properties of
particle may describe a point in momentum space. Let us then consider a frame
in momentum space, which energy and momentum axes, $\pi_{0}$ and $\pi_{1},$
are assumed to be orthogonal. We will call this frame the (momentum)
\textit{photonic frame}. Due to eqs. (\ref{en})\ and (\ref{mom})\ the
dynamical features of particle can be expressed also in terms of reciprocal
space description. In this case we use the (position)\ \textit{photonic frame,
}which orthogonal axes\textit{\ }$\chi_{0}$ and $\chi_{1}$ provide us the
coordinates of \textit{frozen time} and \textit{distance} intervals. We
assume, of course, that energy (time) axis is given in momentum (length)
units, which means that our frame has built-in \textit{comparison scale}
$\eta$.

Let us call the frame for which $\eta=1,$ the \textit{preferred frame}. In
this frame the photon dispersion relation has the form (\ref{en4}) or
(\ref{en44}). Therefore, the \textit{rescaled} dispersion relations
(\ref{en55}), are to be seen as the \textit{photon} dispersion relations given
in frame other then preferred $(\eta\neq1).$ \ The transformations that relate
appropriate \textit{photonic frames} determined for different
\textit{comparison scales, }then\textit{\ }must take respectively the
following forms: in momentum space
\begin{equation}
\left(
\begin{array}
[c]{c}%
\pi_{0}^{\prime}\\
\pi_{1}^{\prime}%
\end{array}
\right)  =\left(
\begin{array}
[c]{cc}%
\frac{1}{\eta} & 0\\
0 & \eta
\end{array}
\right)  \left(
\begin{array}
[c]{c}%
\pi_{0}\\
\pi_{1}%
\end{array}
\right)  ,\label{LM0}%
\end{equation}
and in position space
\begin{equation}
\left(
\begin{array}
[c]{c}%
\chi_{0}^{\prime}\\
\chi_{1}^{\prime}%
\end{array}
\right)  =\left(
\begin{array}
[c]{cc}%
\eta & 0\\
0 & \frac{1}{\eta}%
\end{array}
\right)  \left(
\begin{array}
[c]{c}%
\chi_{0}\\
\chi_{1}%
\end{array}
\right)  .\label{LP0}%
\end{equation}
Additionally, according to (\ref{int1}) and (\ref{int2}) one finds that
transformations (\ref{LM0})\ and (\ref{LP0}) must preserve the condition of
zero invariant interval
\begin{equation}
\left(  \pi_{0}\right)  ^{2}-\left(  \pi_{1}\right)  ^{2}=\left(  \pi
_{0}^{\prime}\right)  ^{2}-\left(  \pi_{1}^{\prime}\right)  ^{2}%
=0,\label{LP10}%
\end{equation}
and
\begin{equation}
\left(  \chi_{0}\right)  ^{2}-\left(  \chi_{1}\right)  ^{2}=\left(  \chi
_{0}^{\prime}\right)  ^{2}-\left(  \chi_{1}^{\prime}\right)  ^{2}%
=0.\label{LP11}%
\end{equation}
The transformations (\ref{LM0}) and (\ref{LP0}), of course, are not unitary.
On the other hand, since the Lorentz transformation (considered in position
space) preserves the condition (\ref{LP11}), it is clear, that it cannot
relate two physically equivalent \textit{photonic frames}. As noticed, in the
case of $\eta\neq1$ the time looses its kinematical meaning and becomes
\textit{frozen} only. Nevertheless, if one neglects the \textit{vital} meaning
of time, on finds that Lorentz transformation (which general form will be
derived later) indeed relates the two \textit{photonic frames} which are
equivalent but in sense of \textit{covariant equivalence}.

\subsection{Description of \textit{photon} state vector in \textit{photonic}
\textit{frame}}

The foregoing discussion shows that \textit{photon} dispersion relation
(\ref{en4}) is form invariant with respect to the \textit{scaling}
\textit{transformation} (\ref{LM0}). The same concern the relation
(\ref{en44}) and transformation (\ref{LP0}). Since the both \textit{scaling}
\textit{transformations} act upon appropriate frame axes, they are the passive
transformations. However, the \textit{scaling transformation} may be
considered also as the active one. The approach requires then that
passive\textit{\ }and active\textit{\ }actions are to be equivalent. But this,
in turn, gives rise to a question about the correct form of vector
representing the \textit{photon }state in \textit{photonic frame}. Let us take
a closer look at this issue.

The two-dimensional energy-momentum \textit{photonic frame}, introduced above,
gives us possibility to represent a physical state in form of two-component
vector $\binom{\Pi_{0}}{\Pi_{1}},$ called further the \textit{bimomentum}$.$
$\ $The action of \textit{active scaling} must then transform $\binom{\Pi_{0}%
}{\Pi_{1}}\rightarrow\binom{\Pi_{0}^{\prime}}{\Pi_{1}^{\prime}}$ where
\begin{equation}
\binom{\Pi_{0}^{\prime}}{\Pi_{1}^{\prime}}=\left(
\begin{array}
[c]{cc}%
\frac{1}{\eta} & 0\\
0 & \eta
\end{array}
\right)  \binom{\Pi_{0}}{\Pi_{1}}.\label{am1}%
\end{equation}
Now, let us assume that \textit{photon} state is represented by
\textit{bimomentum} which components $\Pi_{0}$ and $\Pi_{1}$ are related by
(\ref{en4}). Thus, one would expect that components of new \textit{bimomentum}
$\binom{\Pi_{0}^{\prime}}{\Pi_{1}^{\prime}}$ still fulfill relation
(\ref{en4}). However, since \textit{bimomentum} $\binom{\Pi_{0}^{\prime}}%
{\Pi_{1}^{\prime}}=$ $\binom{\frac{1}{\eta}\Pi_{0}}{\eta\Pi_{1}},$ on finds
that \textit{bimomentum }$\binom{\Pi_{0}^{\prime}}{\Pi_{1}^{\prime}}$ cannot
represent \textit{photon} state in the \textit{photonic frame} in which the
bimomentum $\binom{\Pi_{0}}{\Pi_{1}}$ does. So, one needs to indicate another
\textit{bimomentum} form capable to describe the \textit{photon} state correctly.

As noticed, the covariant description uses the frame which axes, originally
given in different physical units (like energy and momentum), become
``physically equivalent'' after the comparison scale has been imposed. On the
other hand, the magnitudes of energy and momentum of \textit{photon} state
(just in covariant description) are assumed to be equal. This enables
\textit{photon} state vector to be put down in the form \textit{bimomentum}
which one component of two equals zero. Indeed, the scaling transformation
(\ref{LM0})\ acting on state vectors $\binom{\Pi_{0}}{0}$ or $\binom{0}%
{\Pi_{1}},$ does not take them out of this \textit{bimomentum} class. Thus,
the \textit{photonic frame} might be called also the frame which one-component
\textit{bimomenta} describe the \textit{photon} states. The problem related to
ambiguous interpretation of described this way \textit{photon} states, is
discussed next.

\section{The \textit{gauge} frame description}

The analysis has been doing so far is $1+1$ dimensional. However, if one
concentrates only on dynamical properties of single (scalar) particle, such
$1+1$ dimensional analysis turns out to be the \textit{gauge} analysis.
Indeed, let us call the axis along which particle propagates, the
\textit{gauge axis}. So, if the momentum axis covers the \textit{gauge axis,}
one finds that our $1+1$ dimensional energy-momentum frame is to be considered
the \textit{photonic} \textit{gauge frame} in $3+1$ dimensional space.
Consequently, the two-component \textit{bimomentum} $\binom{\Pi_{0}}{\Pi_{1}}$
turns out to be the \textit{gauge }form of four-momentum.

The bimomenta that belong to four different quarters of \textit{photonic}
\textit{gauge frame }correspond to four different kinds of quantum states. In
particular the bimomenta of quarters $I$ ($\Pi_{0}>0,\Pi_{1}>0$) and $II $
($\Pi_{0}>0,\Pi_{1}<0$) describe particle which energies are positive but
momenta are oriented respectively toward the positive and negative direction
of the \textit{gauge axis}. Analogical situation concerns the quarters $III$
($\Pi_{0}<0,\Pi_{1}<0$) and $IV$ ($\Pi_{0}<0,\Pi_{1}>0$) where particle
energies now are negative.

Next, let us consider the \textit{photonic gauge frame} and one-component
\textit{bimomenta} $\binom{\Pi_{0}}{0}$ and $\binom{0}{\Pi_{1}}$ assumed to
represent the \textit{photon} states in this frame. Then, the
\textit{bimomentum} $\binom{\Pi_{0}}{0}$ must describe a \textit{photon} which
energy is positive if $\Pi_{0}>0,$ or negative if $\Pi_{0}<0,$ but the
direction of momentum transfer in both cases is indefinite. Similarly, the
\textit{bimomentum} $\binom{0}{\Pi_{1}}$ must describe a \textit{photon} which
momentum (in given coordinate frame) is positive if $\Pi_{1}>0$ or negative if
$\Pi_{1}<0,$ but energy sign is indefinite now. Discussed in the following the
unitary equivalent representation of \textit{photon bimomenta} allows us to
remove these ambiguities.

\subsection{The \textit{photonic} vs. \textit{Minkowski gauge} frames}

Let us consider the \textit{photonic gauge frame }and a new frame which axes
$p_{0}$ and $p_{1}$ are rotated around the origin of the initial
\textit{photonic} \textit{frame} about $45^{0}$. Then, the relationship
between the coordinates of both frames is given by orthogonal transformation
\begin{equation}
\left(
\begin{array}
[c]{c}%
p_{0}\\
p_{1}%
\end{array}
\right)  =\left(
\begin{array}
[c]{cc}%
\frac{1}{\sqrt{2}} & \frac{1}{\sqrt{2}}\\
\frac{-1}{\sqrt{2}} & \frac{1}{\sqrt{2}}%
\end{array}
\right)  \left(
\begin{array}
[c]{c}%
\pi_{0}\\
\pi_{1}%
\end{array}
\right)  .\label{pm1}%
\end{equation}
The particular choice of rotation angle of $45^{0}$ makes that in the new
frame the coordinates of \textit{photon} vectors are put on diagonals, or, in
other words, on light-cone axes. For example, the \textit{photon} state
described in initial \textit{photonic frame }by \textit{bimomentum }%
$\binom{\Pi_{0}}{0}$\textit{, }and thus having indefinite direction of
momentum\textit{\ }transfer, in the new frame takes the form
\begin{equation}
\left(
\begin{array}
[c]{c}%
P_{0}\\
P_{0}%
\end{array}
\right)  =\left(
\begin{array}
[c]{cc}%
\frac{1}{\sqrt{2}} & \frac{-1}{\sqrt{2}}\\
\frac{1}{\sqrt{2}} & \frac{1}{\sqrt{2}}%
\end{array}
\right)  \left(
\begin{array}
[c]{c}%
\Pi_{0}\\
0
\end{array}
\right)  ,\label{pm11}%
\end{equation}
so that $P_{0}=\Pi_{0}/\sqrt{2}$ and momentum transfer now is well-defined.
Thus, the new frame will be called the \textit{Minkowski gauge frame}. It is
also worthwhile to notice that \textit{\ bimomentum} $\binom{0}{-\Pi_{0}}$ of
\textit{photonic frame}, in the \textit{Minkowski gauge frame} takes the form
\begin{equation}
\left(
\begin{array}
[c]{c}%
P_{0}\\
-P_{0}%
\end{array}
\right)  =\left(
\begin{array}
[c]{cc}%
\frac{1}{\sqrt{2}} & \frac{-1}{\sqrt{2}}\\
\frac{1}{\sqrt{2}} & \frac{1}{\sqrt{2}}%
\end{array}
\right)  \left(
\begin{array}
[c]{c}%
0\\
-\Pi_{0}%
\end{array}
\right) \label{pm12}%
\end{equation}
which describes the \textit{photon} that propagates in opposite direction with
comparison to photon (\ref{pm11}).

Next, let us see how the \textit{scaling transformation} of \textit{photonic}
frame acts on\textit{\ Minkowski gauge frame} axes.

\subsection{Scaling transformation of \textit{Minkowski gauge} frame}

Let us consider again the \textit{scaling transformation} of \textit{photonic
frame} (\ref{LM0}). If one considers the two \textit{photonic frames} defined
for two different \textit{comparison scales} $\eta_{1}$ and $\eta_{2},$ one
finds that \textit{scaling transformation} relating these two frames still has
the same form (\ref{LM0}), but now $\eta=\eta_{1}\cdot\eta_{2}.$ Thus, the
form of transformation (\ref{LM0}) is general. Since there is unitary
equivalence of \textit{photonic} and \textit{Minkowski gauge} frames it is
advisable to determine the transformation of \textit{Minkowski frame} induced
by the \textit{scaling transformation} of \textit{photonic frame}. For that
purpose, let us consider the two \textit{photonic gauge frames} given for two
different \textit{comparison scales} $\eta_{1}$ and $\eta_{2},$ and related to
them the two \textit{Minkowski gauge frames. }The corresponding
transformations that relate both kinds of frames provide the formulas
(\ref{pm1}) and
\begin{equation}
\left(
\begin{array}
[c]{c}%
p_{0}^{\prime}\\
p_{1}^{\prime}%
\end{array}
\right)  =\left(
\begin{array}
[c]{cc}%
\frac{1}{\sqrt{2}} & \frac{1}{\sqrt{2}}\\
\frac{1}{-\sqrt{2}} & \frac{1}{\sqrt{2}}%
\end{array}
\right)  \left(
\begin{array}
[c]{c}%
\pi_{0}^{\prime}\\
\pi_{1}^{\prime}%
\end{array}
\right)  .\label{pm2}%
\end{equation}
Thus, according to (\ref{LM0}), (\ref{pm1})\ and (\ref{pm2}) one obtains
\begin{align}
\left(
\begin{array}
[c]{c}%
p_{0}^{\prime}\\
p_{1}^{\prime}%
\end{array}
\right)   & =\left(
\begin{array}
[c]{cc}%
\frac{1}{\sqrt{2}} & \frac{1}{\sqrt{2}}\\
\frac{1}{-\sqrt{2}} & \frac{1}{\sqrt{2}}%
\end{array}
\right)  \left(
\begin{array}
[c]{cc}%
\frac{1}{\eta} & 0\\
0 & \eta
\end{array}
\right)  \left(
\begin{array}
[c]{c}%
\pi_{1}\\
\pi_{0}%
\end{array}
\right) \nonumber\\
& =\left(
\begin{array}
[c]{cc}%
\frac{1}{\sqrt{2}} & \frac{1}{\sqrt{2}}\\
\frac{1}{-\sqrt{2}} & \frac{1}{\sqrt{2}}%
\end{array}
\right)  \left(
\begin{array}
[c]{cc}%
\frac{1}{\eta} & 0\\
0 & \eta
\end{array}
\right)  \left(
\begin{array}
[c]{cc}%
\frac{1}{\sqrt{2}} & \frac{-1}{\sqrt{2}}\\
\frac{1}{\sqrt{2}} & \frac{1}{\sqrt{2}}%
\end{array}
\right)  \left(
\begin{array}
[c]{c}%
p_{0}\\
p_{1}%
\end{array}
\right)  ,
\end{align}
which finally leads to the known expression
\begin{equation}
\left(
\begin{array}
[c]{c}%
p_{0}^{\prime}\\
p_{1}^{\prime}%
\end{array}
\right)  =\left(
\begin{array}
[c]{cc}%
cosh\xi &  sinh\xi\\
sinh\xi &  cosh\xi
\end{array}
\right)  \left(
\begin{array}
[c]{c}%
p_{0}\\
p_{1}%
\end{array}
\right)  ,\label{pm4}%
\end{equation}
where $\xi=ln\eta$ and $\eta=\eta_{1}\cdot\eta_{2}.$ Eq. (\ref{pm4}), of
course, is the Lorentz transformation of $1+1$ dimensional momentum
\textit{gauge frame}. One easily finds that Lorentz transformation (\ref{pm4})
is equivalent to the \textit{scaling transformation} of light-cone axes in
\textit{Minkowski gauge frame}, or alternatively, to the \textit{scaling
transformation} of axes of \textit{photonic frame}. This shows again (cf. eqs.
(\ref{LP10}), (\ref{LP11})) that source of Lorentz covariance is the
\textit{freedom of choice of comparison scale, }which, as it comes from above,
is embedded already at the level of quantum-mechanical description.

The issue of \textit{photon} dispersion relation made visible that concept of
\textit{comparison scale }may be used to combine\textit{\ }the postulates of
Planck and de Broglie in Lorentz covariant way. The following section in a
natural way extends above analysis into description of relativistic particle
with mass.

\section{Composite structure of massive states}

Actually, there are two well-known examples of theoretical approaches where
quantum objects are treated as extended ones. The first one provides the
string theory \cite{string}. The other, much more elementary, however much
more closely related to usual quantum mechanics, provides the model of
covariant harmonic oscillator \cite{CO1},\cite{K1}. Nevertheless, these two
quite different approaches share the same idea of particle mass introduction.
Indeed, because of assumed particle internal space-time structure, in both
cases particle mass does not appear as a description parameter (like in the
case of know Dirac equation) but, one may briefly say, it is extracted from
massless wave equation. Presented below simple quantum-mechanical analysis in
some sense confirm these results and indicates that massive states, indeed,
can be constructed with aid of the massless ones.

The transformation (\ref{pm1}) representing the transition from the
\textit{photonic} to \textit{Minkowski gauge frame} was introduced to describe
the \textit{photon} states in clear-cut way. This transformation, however,
also enable us to identify a \textit{two-photon} state represented by
two-component \textit{bimomentum} of \textit{photonic frame} with a massive
state represented by \textit{bimomentum }of \textit{Minkowski gauge frame}.
Indeed, let us consider the \textit{bimomentum}
\begin{equation}
\binom{\Pi_{0}}{\Pi_{1}}=\binom{\frac{mc}{\sqrt{2}}}{-\frac{mc}{\sqrt{2}}%
},\label{m1}%
\end{equation}
assumed to describe a particle with mass $m$ in \textit{photonic gauge frame}.
According to (\ref{pm1}), the corresponding form of \textit{bimomentum}
(\ref{m1}) in \textit{Minkowski gauge frame} is
\begin{equation}
\binom{p_{0}}{p_{1}}=\binom{mc}{0},\label{m2}%
\end{equation}
which basically describes particle at rest. So, one finds that
\textit{bimomentum} of particle at rest in \textit{Minkowski gauge frame} is
unitary equivalent to the \textit{bimomentum} of \textit{two-photon} state,
where each of the photons in Minkowski frame must propagate in opposite
directions. Note, that such combination of \textit{photon} states suggests
that effectively introduced massive state is to be considered rather extended
then point-like.

Next, let us consider the transformation (\ref{pm4}) and note that it may be
understood: (1) as the \textit{passive} one, when both \textit{bimomenta
}$\binom{p_{0}}{p_{1}}$ and $\binom{p_{0}^{\prime}}{p_{1}^{\prime}}$ refer to
the same physical state but they are described in two different reference
frames (i.e. in frames based on different \textit{comparison scales}), or (2)
as the \textit{active} one when both \textit{bimomenta} are considered in the
same \textit{preferred frame} but refer to two different physical states. In
the case (1) $\eta$ is the \textit{passive} parameter, whereas in the case (2)
is the \textit{active }one\textit{.} So, in the latter case the $\eta
-parametrization$ does not concern the frame characteristics but describes
dynamical features of the state. We write this down in the explicit form
\begin{equation}
\left(
\begin{array}
[c]{c}%
p_{0}\\
p_{1}%
\end{array}
\right)  =\left(
\begin{array}
[c]{cc}%
\gamma & \gamma\beta\\
\gamma\beta & \gamma
\end{array}
\right)  \left(
\begin{array}
[c]{c}%
mc\\
0
\end{array}
\right)  ,\label{Lor2}%
\end{equation}
where $\gamma=cosh\xi$ and $\gamma\cdot\beta=sinh\xi$. Formally, the
transformation matrixes given in (\ref{pm4})\ and (\ref{Lor2})\ are identical.
However, the parameters $\gamma$ and $\beta$ (which kinematical meaning will
be discussed latter) were introduced to emphasis that their values need to be
refer to the frame set for $\eta=1.$

Note, that massive states, which are assumed to be physically observed, now
emerge as the effective ones in Minkowski frame description. Indeed, currently
discussed mass-shell state was made up of two \textit{photon} states.
Furthermore, since the momentum transfer of these two \textit{photon} states
is opposite, the suspicion that massive particles should have some space-time
extensions become justified. On the other hand, since the momentum transfer
for each interaction process is assumed to be local, the picture of
point-particle maintains its validity too. But this in turn means that
classical picture of point particle and the quantum one of extended quantum
object not necessarily have to be mutually exclusive.

\section{Quantum-mechanical foundations of Minkowski space-time}

In this section we extend $1+1$ dimensional \textit{gauge} analysis onto
entire $3+1$ dimensional space-time. We show that the origin of symmetry based
on homogenous Lorentz group\ is still the same quantum-mechanical
\textit{freedom of choice of comparison scale.} Indeed, the scaling parameter
$\eta$ turns out to be the only relevant parameter of six-parameter Lorentz
group, which makes sharp distinction between relativistic and classical
approach. We show this explicitly by deriving the diagonal form of the
Lorentz-boost transformation matrix. In first subsection we indicate that
Minkowski space-time, first of all, is to be consider energy-momentum
reciprocal space, what means that space-time description, in general,
\ provides only the \textit{frozen time }meaning\textit{. }The next two
following subsections are devoted to issue of kinematical meaning of time in
relativistic approach.

\subsection{Diagonal form of Lorentz-boost transformation}

The \textit{scaling transformation} expressing the \textit{freedom of choice
of comparison scale} of $1+1$ dimensional \textit{photonic} \textit{frame} in
momentum space was shown to have the form (\ref{LM0}). Similarly it was shown
that eq. (\ref{LP0}) describes the space-time transformation equivalent to
(\ref{LM0}) but considered in reciprocal momentum space. Now, we generalize
our analysis onto entire space, assuming that the \textit{gauge-axis }\ of
$1+1$ dimensional \textit{photonic frame} is the $x-axis$ of\textit{\ }$3+1$
dimensional frame. In this frame the \textit{scaling (gauge) transformation
}(\ref{LP0})\ must take the form
\begin{equation}
\left(
\begin{array}
[c]{c}%
\chi_{0}^{\prime}\\
\chi_{1}^{\prime}\\
\chi_{2}^{\prime}\\
\chi_{3}^{\prime}%
\end{array}
\right)  =\left(
\begin{array}
[c]{cccc}%
\eta & 0 & 0 & 0\\
0 & \frac{1}{\eta} & 0 & 0\\
0 & 0 & 1 & 0\\
0 & 0 & 0 & 1
\end{array}
\right)  \left(
\begin{array}
[c]{c}%
\chi_{0}\\
\chi_{1}\\
\chi_{2}\\
\chi_{3}%
\end{array}
\right)  .\label{LP1}%
\end{equation}
According to (\ref{pm1}), one finds that appropriate transitions from the
\textit{photonic} to \textit{Minkowski gauge} frame in the case of $3+1$
dimensional space-time must be given by orthogonal transformations
\begin{equation}
\left(
\begin{array}
[c]{c}%
x_{0}\\
x_{1}\\
x_{2}\\
x_{3}%
\end{array}
\right)  =\left(
\begin{array}
[c]{cccc}%
\frac{1}{\sqrt{2}} & \frac{1}{\sqrt{2}} & 0 & 0\\
\frac{-1}{\sqrt{2}} & \frac{1}{\sqrt{2}} & 0 & 0\\
0 & 0 & 1 & 0\\
0 & 0 & 0 & 1
\end{array}
\right)  \left(
\begin{array}
[c]{c}%
\chi_{0}\\
\chi_{1}\\
\chi_{2}\\
\chi_{3}%
\end{array}
\right)  ,\label{LP2}%
\end{equation}
and
\begin{equation}
\left(
\begin{array}
[c]{c}%
x_{0}^{\prime}\\
x_{1}^{\prime}\\
x_{2}^{\prime}\\
x_{3}^{\prime}%
\end{array}
\right)  =\left(
\begin{array}
[c]{cccc}%
\frac{1}{\sqrt{2}} & \frac{1}{\sqrt{2}} & 0 & 0\\
\frac{-1}{\sqrt{2}} & \frac{1}{\sqrt{2}} & 0 & 0\\
0 & 0 & 1 & 0\\
0 & 0 & 0 & 1
\end{array}
\right)  \left(
\begin{array}
[c]{c}%
\chi_{0}^{\prime}\\
\chi_{1}^{\prime}\\
\chi_{2}^{\prime}\\
\chi_{3}^{\prime}%
\end{array}
\right)  .\label{LP3}%
\end{equation}
Thus, combining of eqs. (\ref{LP1}), (\ref{LP2}) and (\ref{LP3}) one finds
that \textit{scaling transformation} of \textit{photonic gauge frame}
corresponds to the Lorentz transformation of $3+1$ dimensional
frame\textit{\ }
\begin{equation}
\left(
\begin{array}
[c]{c}%
x_{0}^{\prime}\\
x_{1}^{\prime}\\
x_{2}^{\prime}\\
x_{3}^{\prime}%
\end{array}
\right)  =\left(
\begin{array}
[c]{cccc}%
cosh\xi & -sinh\xi & 0 & 0\\
-sinh\xi &  cosh\xi & 0 & 0\\
0 & 0 & 1 & 0\\
0 & 0 & 0 & 1
\end{array}
\right)  \left(
\begin{array}
[c]{c}%
x_{0}\\
x_{1}\\
x_{2}\\
x_{3}%
\end{array}
\right)  ,\label{Lor1}%
\end{equation}
where\ $\xi=ln\eta$. In matrix notation eq. (\ref{Lor1}) is
\begin{equation}
x^{\prime}=A_{0}x.\label{Lor3s}%
\end{equation}

Next, let us determine the new form of transformation matrix $A_{0}$ in the
case, where both (mutually parallel) Minkowski frames are additionally rotated
about their origins in the same way. This corresponds to the situation, where
the \textit{gauge-direction} does not match any of Minkowski frame space axes.
Let then us rotate the initial frames, first (let say) in $xy$ and next in
$yz$ plane, according to
\begin{equation}
R_{yz}R_{xy}x^{\prime}=R_{yz}R_{xy}A_{0}R_{xy}^{-1}R_{yz}^{-1}R_{yz}%
R_{xy}x,\label{Lor3r}%
\end{equation}
where
\begin{equation}
R_{xy}=\left(
\begin{array}
[c]{cccc}%
1 & 0 & 0 & 0\\
0 & cos\varphi & -sin\varphi & 0\\
0 & sin\varphi &  cos\varphi & 0\\
0 & 0 & 0 & 1
\end{array}
\right)  ,\text{ }R_{yz}=\left(
\begin{array}
[c]{cccc}%
1 & 0 & 0 & 0\\
0 & 1 & 0 & 0\\
0 & 0 & cos\psi & -sin\psi\\
0 & 0 & sin\psi &  cos\psi
\end{array}
\right)  ,\label{rotM}%
\end{equation}
so that, the angles $\varphi$ and $\psi$ describe the rotations around the $z$
and $x$ axes respectively. The explicit form of eq. (\ref{Lor3r}) is given by
\begin{equation}
\left(
\begin{array}
[c]{c}%
x_{0}^{\prime}\\
x_{1}^{\prime}\\
x_{2}^{\prime}\\
x_{3}^{\prime}%
\end{array}
\right)  =\left(
\begin{array}
[c]{cccc}%
\gamma & -\gamma\beta_{1} & -\gamma\beta_{2} & -\gamma\beta_{3}\\
-\gamma\beta_{1} & 1+\frac{(\gamma-1)}{\beta^{2}}\beta_{1}^{2} & \frac
{(\gamma-1)}{\beta^{2}}\beta_{1}\beta_{2} & \frac{(\gamma-1)}{\beta^{2}}%
\beta_{1}\beta_{3}\\
-\gamma\beta_{2} & \frac{(\gamma-1)}{\beta^{2}}\beta_{1}\beta_{2} &
1+\frac{(\gamma-1)}{\beta^{2}}\beta_{2}^{2} & \frac{(\gamma-1)}{\beta^{2}%
}\beta_{2}\beta_{3}\\
-\gamma\beta_{3} & \frac{(\gamma-1)}{\beta^{2}}\beta_{1}\beta_{3} &
\frac{(\gamma-1)}{\beta^{2}}\beta_{2}\beta_{3} & 1+\frac{(\gamma-1)}{\beta
^{2}}\beta_{3}^{2}%
\end{array}
\right)  \left(
\begin{array}
[c]{c}%
x_{0}\\
x_{1}\\
x_{2}\\
x_{3}%
\end{array}
\right)  ,\label{Lor4}%
\end{equation}
where $x$ and $x^{\prime}$ are now the ``rotated equivalents'' of those of eq.
(\ref{Lor3s}) and $\beta_{1}/\beta=cos\varphi$, $\beta_{2}/\beta
=sin\varphi\cdot cos\psi,$ $\beta_{3}/\beta=sin\varphi\cdot sin\psi$, and thus
$\mathbf{\beta=}(\beta_{1},\beta_{2},\beta_{3}).$ Eq. (\ref{Lor4}) written in
short cut form is
\begin{equation}
x^{\prime}=\mathbf{A}\text{ }x.\label{Lor4s}%
\end{equation}
To recollect, the matrix $\mathbf{A}$ represents the \textit{Lorentz-boost}
transformation. Traditionally it is derived by means of boost generators of
the Lorentz group and the Taylor expansion \cite{Jack}. The way in which it
was constructed now, allows us to express the matrix $\mathbf{A}$ in the form
\begin{equation}
\mathbf{A}=\mathbf{RU\Lambda}_{\eta}\mathbf{U}^{-1}\mathbf{R}^{-1}%
,\label{Lor4A}%
\end{equation}
where
\begin{equation}
\mathbf{\Lambda}_{\eta}=\left(
\begin{array}
[c]{cccc}%
\eta & 0 & 0 & 0\\
0 & \frac{1}{\eta} & 0 & 0\\
0 & 0 & 1 & 0\\
0 & 0 & 0 & 1
\end{array}
\right)  ,\text{ }\mathbf{U}=\left(
\begin{array}
[c]{cccc}%
\frac{1}{\sqrt{2}} & \frac{1}{\sqrt{2}} & 0 & 0\\
\frac{-1}{\sqrt{2}} & \frac{1}{\sqrt{2}} & 0 & 0\\
0 & 0 & 1 & 0\\
0 & 0 & 0 & 1
\end{array}
\right)  ,\text{ }\mathbf{R}=R_{yz}\circ R_{xy}.\label{rotA}%
\end{equation}
Since neither of the matrixes $\mathbf{R}$ nor $\mathbf{U}$ is singular, the
diagonal form of the matrix $\mathbf{A}$ is given by $\mathbf{\Lambda}_{\eta}%
$. This shows that three boost parameters $\beta_{1},\beta_{2}$ and $\beta
_{3}$ emerge as a result of \ $\eta$-factor ``splitting'' induced by the
orientation of the \textit{gauge direction} in three dimensional real-space.
Additional consideration of Euclidean real-space rotations, of course, is of
no importance from the point of view of current analysis.

Let us briefly summarize the above results. It has been shown that idea of
\textit{comparison scale} implanted on very quantum-mechanical ground in
straightforward manner leads to the concept of homogenous Lorentz group. The
peculiarity of Lorentz symmetry comes from the fact that it embraces the
features of geometrical and dynamical description (but not just only
geometrical, as it is commonly thought). The Lorentz transformation emerge
then quite naturally as the passive transformation of space-time, where the
space-time turns out to be the reciprocal energy-momentum space. Therefore at
the level of Lorentz covariant notation the time might have only the
\textit{frozen} meaning. On the other hand the quantum basis of Lorentz
symmetry appears in the obvious way.

\subsection{Kinematical meaning of Lorentz formulas}

The Lorentz transformation formulas, due to textbook interpretation, involves
the velocity as purely relative quantity. Indeed, if two, so-called, inertial
observers (placed in their own rest frames) are in relative motion at velocity
$w,$ each of them is expected to measure the same velocity $w$ of his moving
co-partner. This expectation, however, is based on pure classical assumption
that both observers, indeed, are quite equivalent. However, this is not the
case of quantum measurement where one of the observers is to be substituted
for a quantum object. Furthermore, in contrary to classical measurement, as
already noticed, one cannot neglect the influence of measuring apparatus on
the final results. In other words, the quantum observation always is the
\textit{preferred frame} observation. Thus, in particular, within the
framework quantum measurement, the notion of (particle) velocity cannot be
considered relative. If quantum description involves the Lorentz symmetry
(what has been clearly shown already) then the orthodox point of view at the
kinematical meaning of Lorentz formulas needs a thorough refinement. We start
to discuss this issue by considering another velocity aspect, namely, its
relationship with particle energy and momentum.

According to (\ref{Lor2}) the elements of transformation matrix (\ref{pm4})
may be expressed by means of parameters
\begin{equation}
\gamma=cosh\xi\text{ \ \ \ and \ \ }\gamma\cdot\beta=sinh\xi.\label{ptex}%
\end{equation}
On the other hand the textbook formulas:
\begin{equation}
\gamma=\frac{1}{\sqrt{1-w^{2}/c^{2}}}\text{ and \ }\beta=\frac{w}%
{c},\label{wvelo}%
\end{equation}
tell us how to relate the parameters (\ref{ptex}) with some velocity $w,$
interpreted as the velocity of point-particle moving at the observer rest
frame. According to (\ref{wvelo}) and (\ref{Lor2}) the particle energy and
momentum expressed by means of velocity $w$ are
\begin{equation}
E=\frac{mc^{2}}{\sqrt{1-w^{2}/c^{2}}},\text{ \ \ \ }p=\frac{mw}{\sqrt
{1-w^{2}/c^{2}}}.\label{Epfor}%
\end{equation}
In similar manner the Lorentz transformation (\ref{Lor3s}) combined with
expressions (\ref{ptex}) and (\ref{wvelo}) takes the well-known form
\begin{equation}
\text{\ }t^{\prime}=\frac{t-(w/c^{2})\cdot x}{\sqrt{1-w^{2}/c^{2}}},\text{
\ }x^{\prime}=\frac{x-w\cdot t}{\sqrt{1-w^{2}/c^{2}}},\text{ }y^{\prime
}=y,\text{ \ }z^{\prime}=z.\label{LorF1}%
\end{equation}
Since the velocity $w$ is the kinematical parameter of observer rest frame,
the same must concern the time $t$ and distance $x.$ However, the key issue is
that due to the orthodox interpretation, $t^{\prime}$ and $x^{\prime}$ are
also kinematical parameters but referring to the particle rest frame. The
arguments given through the paper clearly indicate that currently such
interpretation is not allowed. In particular, it is commonly thought that the
time passing at the rest frames of observer and particle are different. The
source of such misleading interpretation, let us emphasize once more, is a
false conviction that the picture of particle moving at observer rest frame is
physically equivalent to the one where the particle stay at rest but the
observer moves. There is no equivalence between the rest frames of observer
and particle. Furthermore, in fact, it is even hard to say what does the
notion of particle rest frame mean. The particle is only a quantum object and
cannot be identified with any physical observer. On the other hand, it does
make sense to say that the observer, in his own rest frame, observers
different quantum states of given quantum object. This is why the Lorentz
transformation cannot preserve the \textit{vital time} meaning, and thus the
formulas (\ref{LorF1}) cannot relate two physically equivalent frames.

\subsection{Quantum-mechanical explanation of time dilatation effect}

Presumably, the most natural question arising now is how to explain the,
so-called, time dilatation effect. Although satisfactory discussion of this
issue requires a field theory approach, as well as, a new insight into
Heisenberg uncertainty principles \cite{PK1}, a simply quantum-mechanical
explanation can be given right now.

First, let us note that the parametrization of $\gamma$ and $\beta$
\ (\ref{wvelo}) is not unique. An alternative expressions can be found by
introducing a new velocity $v$ related to $w$ according to
\begin{equation}
v=\frac{w}{\sqrt{1-w^{2}/c^{2}}}.\label{vw}%
\end{equation}
This yields
\begin{equation}
\gamma=\sqrt{1+v^{2}/c^{2}}\text{ \ and \ }\gamma\cdot\beta=\frac{v}%
{c}.\label{vw2}%
\end{equation}
As a result, the energy and momentum formulas (\ref{Epfor}) may be replaced by
the new ones
\begin{equation}
E=mc^{2}\sqrt{1+v^{2}/c^{2}},\text{ \ \ }p=mv.\label{Epfor2}%
\end{equation}
One sees that momentum (\ref{Epfor2}) has the classical form, whereas the
energy is regular function of $v$ in the whole range. Since the velocities
which magnitudes are grater then $c$ are, in general, not observed, the
physical interpretation of expressions (\ref{Epfor2}) is rather troublesome.
The most straightforward way to overcome this difficulty is to treat both
velocities $w$ and $v$ on equal physical footing. Indeed, let us assume again
that particle seen as a quantum object may exist under the cover of two forms,
seemingly quite different, but in fact, complement one another.

The first form is assumed to have point-particle features, whilst the other
the features of extended quantum object. Let us then assume that in the first,
classical-like case, the moving particle is identified with point-like object
of mass $m,$ which covers the distance $\Delta l$ at velocity $v$ in time
period $\Delta t_{v}.$

Next, to ascribe to the particle of the second possible form of existence
(which does not have classical counterpart), let us make a striking assumption
that this extended quantum form correspond to the state which is temporarily
localized in observer rest frame, which longitudinal extension is just $\Delta
l$ and which life-time is $\Delta t_{w}$ where
\begin{equation}
\Delta l=w\Delta t_{w}=v\Delta t_{v}.\label{dist}%
\end{equation}
According to (\ref{vw})\ one finds that relation between the time intervals
$\Delta t_{w}$ and $\Delta t_{v}$ is
\begin{equation}
\Delta t_{w}=\frac{\Delta t_{v}}{\sqrt{1-w^{2}/c^{2}}}.\label{dila}%
\end{equation}
Thus, the assumption of quantum (localized and extended) and classical
(point-like and not localized) particle nature provide us simply and plausible
interpretation of formulas (\ref{dist}) and (\ref{dila}). In particular, on
finds that formula (\ref{dila}) relates the life-times of moving
point-particle and corresponding quantum state. If $\Delta t_{v}$ is the life
time of unstable particle, then simple quantum-mechanical approach may explain
why ``relativistic particle'' produced at some point $x_{A}$ can be found at
point $x_{B}$ distant form $x_{A}$ much more than the light pulse can travel
at the time period $\Delta t_{v}.$ On the other hand, if the life time of
assumed temporarily localized state is also the time needed for the
measurement to establish the point-particle position, one finds that causality
is always preserved.

The complete quantum description of extended quantum state (now introduced
only in classical-like manner), of course, should be characterized through the
covariant probability distribution, localized in some space-time region.
Currently, let us stress only that examples of such distributions provide us
just mentioned solutions of covariant harmonic oscillator. To maintain the
paper simplicity, however, it is advisable to separate this issue from the
discussion right now.

Finally, let us note that Lorentz transformation formulas (\ref{LorF1})\ can
be write down in alternative form involving velocity $v$ instead of $w$,
namely
\begin{equation}
t^{\prime}=t\sqrt{1+v^{2}/c^{2}}-\frac{v}{c^{2}}x,\text{ \ \ }x^{\prime
}=x\sqrt{1+v^{2}/c^{2}}-vt,\text{ \ \ }y^{\prime}=y,\text{\ \ }z^{\prime
}=z.\label{LorF2}%
\end{equation}
The formulas (\ref{LorF1}) and (\ref{LorF2}) express the same \textit{passive}
transformation (\ref{Lor1}), which corresponds to the same change of
comparison $\eta-scale.$ Thus, the kinematical meaning of $t^{\prime}$ and
$x^{\prime}$ in formulas (\ref{LorF1}) and (\ref{LorF2}) is lost. The case of
$\eta\approx1,$ corresponding to small velocity values, reduces the both kinds
of Lorentz formulas to the Galilean form, and thus provides us the ``true''
classical limit.

\section{Concluding remarks}

Discovery and research of classical electromagnetic field, undoubtedly, occupy
central position in development of physics. Although experimental basis of
Maxwell equations is thought to be the classical one, in fact, the very object
of Maxwell equations, as well as, their algebraic structure go much beyond the
framework of pure Newton world picture. One of the main reason for that is, of
course, that Newton theory deals with the concept of material point, which, in
wave description, does not have a simple counterpart. The most closely related
to this notion is the one of quasiparticle, but it appears already at the
level of second quantization. Furthermore, the mechanics of material point(s)
and waves, in both cases called the classical, start with descriptions based
on different symmetry groups. Thus, an arising question is whether the Lorentz
symmetry might be called classical (i.e. not quantum) at all. Note that simple
physical observations, such as the blackbody radiation or the photoelectric
effect, clearly revel the quantum nature of electromagnetic field and confirm
the validity of \textit{photon dispersion relation} (\ref{en4}). However, as
it was explained, from the quantum-mechanical point of view, such dispersion
relation can be set only with accuracy to the \textit{freedom of choice of
comparison scale}. Thus, the practical measurements have to use the physical
\textit{photon system} as the preferred or standard one in order to plot the
values of energies and momenta of other quasiparticle excitations on absolute,
i.e. \textit{preferred frame,} energy and momentum scale. Here, one needs to
emphasize that such distinguished role of \textit{photon system} has been
already noticed in \cite{Amelino1}. Nevertheless, the Lorentz symmetry always
is thought to manifest the classical, i.e. purely geometrical properties of space-time.

Indeed, the notion of space-time takes its origin from the classical (i.e.
Newton-like) analysis of electromagnetic field. The six-parameter, homogenous
Lorentz group, or in more general case (when space-time translations are
included), the ten-parameter Poincar\'{e} group, determine the algebraic
properties of space-time. According to orthodox view the physical background
of this symmetry is Einstein idea of relativity. It is well-know that such
point of view is the source of old twin-paradox. Although most of the
physicist get used to it and see noting wrong with that, is worthwhile to
notice that this paradox still comes back to life, however, under the new
cover \cite{Amelino1}, \cite{DSR}. Then, one may ask what forces us to believe
something which is to be challenged. A few sentences taken from the book of
Kim and Noz \cite{K1} might be the answer to this question.

``Developing a new physical theory usually requires a new set of mathematical
formulas. There are in general two different approaches to this problem.
According to Eddington we have to understand all the physical principles
before writing down the first mathematical formula. According to Dirac,
however, it is more profitable to construct plausible mathematical devices
which can describe quantitatively the real world, and then add physical
interpretation to the mathematical formalism. Both special relativity and
quantum mechanics were developed in Dirac's way, and most of the new physical
models these days are developed in this way''.

This paper, in the most elementary way, indicates that quantum mechanics
itself provides plausible and reasonable translation of special relativity
language. As noticed by Kim: ``If not possible, it is very difficult to
formulate Lorentz boosts for rigid bodies. On the other hand, it seems to be
feasible to boost waves'' \cite{K0}. The view that Lorentz symmetry reflects
the relativity of time and length measure is a misunderstanding. To justify
the introduction of relativistic description the support of orthodox
relativity principle is not needed. Note that derivation of Lorentz
transformation (\ref{pm4}) or (\ref{Lor1}), in fact, encompasses only the
quantum-mechanical postulates of Planck and de Broglie and the \textit{idea of
comparison scale}. Thus, one easily finds that Lorentz space-time
transformation is simply the passive transformation of reciprocal
four-momentum space. The time that undergoes relativistic transformations
rules is only the \textit{frozen time}, whereas the kinematical
(\textit{vital}) time meaning can be ascribed only to the\textit{\ preferred
frame,} which is the rest frame of the observer.

In summary, the main conclusion of this paper is that Lorentz symmetry is the
symmetry of quantum description resulting from the \textit{freedom of choice
of comparison scale}. The Lorentz symmetry then is the symmetry of
\textit{preferred frame }description.

\end{document}